\documentclass[twocolumn,showpacs,preprintnumbers]{revtex4}%
\usepackage{amsfonts}
\usepackage{amsmath,epsfig}
\usepackage{graphicx}
\usepackage{dcolumn}
\usepackage{bm}
\usepackage{amssymb}
\usepackage{amsmath}%
\begin{document}
\title{Can very compact and very massive neutron stars both exist?}
\author{Alessandro Drago$^{\text{(a)}}$, Andrea Lavagno$^{\text{(b)}}$ and Giuseppe Pagliara$^{\text{(a)}}$}
\affiliation{$^{\text{(a)}}$Dip.~di Fisica e Scienze della Terra dell'Universit\`a di Ferrara and INFN
Sez.~di
Ferrara, Via Saragat 1, I-44100 Ferrara, Italy}
\affiliation{$^{\text{(b)}}$ Department of Applied Science and Technology, Politecnico
di Torino, Italy
Istituto Nazionale di Fisica Nucleare (INFN), Sezione di Torino, Italy}

\begin{abstract}
The existence of neutron stars with masses of $\sim 2\,M_{\odot}$
requires a stiff equation of state at high densities. On the other
hand, the necessary appearance also at high densities of new degrees
of freedom, such as hyperons and $\Delta$ resonances, can lead to a
strong softening of the equation of state with resulting maximum
masses of $\sim 1.5\, M_{\odot}$ and radii smaller than $\sim 10$
km. Hints for the existence of compact stellar objects with very small
radii have been found in recent statistical analysis of quiescent
low-mass X-ray binaries in globular clusters. We propose an
interpretation of these two apparently contradicting measurements,
large masses and small radii, in terms of two separate families of
compact stars: hadronic stars, whose equation of state is soft, can be
very compact, while quark stars, whose equation of state is stiff, can
be very massive.  In this respect an early appearance of $\Delta$
resonances is crucial to guarantee the stability of the branch of
hadronic stars.  Our proposal could be tested by measurements of radii
with an error of $\sim 1$ km, which is within reach of the planned
LOFT satellite, and it would be further strengthened by the discovery
of compact stars heavier than $\sim 2\,M_{\odot}$.
\end{abstract}

\maketitle

The recent discovery of Compact Stars (CSs) having a mass of the order
of $2\,M_{\odot}$ \cite{Demorest:2010bx,Antoniadis:2013pzd} puts
rather severe constraints on the Equation of State (EoS) of matter at
large densities. It is clear that matter inside a compact star,
i.e. $\beta$-stable and charge neutral matter, has to be stiff to
allow such massive configurations. On the other hand, we know that by
increasing the density new degrees of freedom come into the game, for
instance hyperons and maybe deconfined quarks. These new ingredients
soften the EoS close to their production threshold, but by introducing
repulsive interactions the EoS can be stiff enough at large densities
to support a $2\,M_{\odot}$ configuration. Examples of hyperonic stars
\cite{Bednarek:2011gd,Weissenborn:2011ut} and of hybrid stars
\cite{Weissenborn:2011qu,Bonanno:2011ch,Zdunik:2012dj,Klahn:2013kga}
satisfying that constraint exist in the literature, although special
limits on the parameters' values have to be imposed.  Also Quark
  Stars (QSs), stellar objects composed entirely by quark matter
  (which could exist if the so called Bodmer-Witten hypothesis holds
  true) \cite{Bodmer:1971we,1984PhRvD..30..272W,Farhi:1984qu}, can
  satisfy the constraint \cite{Alford:2006vz,Weissenborn:2011qu}.
  It is however unlikely that all CSs are QSs: the latter are
  probably unable to exhibit glitches \cite{Madsen:1989pg} \footnote{The crystalline LOFF quark phase could possibly 
solve this problem \cite{Anglani:2013gfu} although much theoretical work still remains in order to construct the 
vortices in the crystalline phase
and to calculate their pinning force.} 
and to explain
the data on quasi-periodic oscillations \cite{Watts:2006hk}. It is
therefore clear that, while the $2\,M_{\odot}$ limit allows for
exclusion of entire classes of EoSs which are just too soft, by itself
it is not able to single out the EoS of matter at large densities.

A way to strongly reduce the uncertainty on the EoS would be to
measure the radius of a few CSs, but unfortunately the precise
measurement of the radius has up to now proved to be extremely
difficult, since it is in most cases based on specific assumptions
concerning e.g. the atmosphere and the distance of the object under
investigation. Different analyses often lead to opposite
conclusions. There have been therefore claims of very small radii, of
the order or smaller than about 10 km \cite{Guillot:2013wu}, while
other analyses suggest for the same objects significantly larger
radii, of the order of 12 km \cite{Lattimer:2013hma}.  It is clear
that a precise and model independent measurement of the radius of at
least a few CSs is crucial to finally provide the necessary
information which will allow the extraction of the EoS of stellar
matter at large densities. New satellites have been proposed and in
particular LOFT \cite{Feroci:2011jc,Feroci:2012qh} claims to be able
to measure the radius of a CS, in a few cases, with a precision of the
order of 1 km, small enough to distinguish between the two
possibilities discussed above.

From the theoretical side the families of nucleonic
\cite{Akmal:1998cf}, hyperonic \cite{Weissenborn:2011ut} and hybrid
stars \cite{Zdunik:2012dj,Alford:2013aca,Kurkela:2010yk}, 
stiff enough to reach $2\,M_{\odot}$, all
provide radii which are not too small, typically larger than about
11.5-12 km for the canonical $1.4 \,M_{\odot}$ star.  In studies based
on piecewise polytropic extensions of EoSs derived within chiral
effective field theory up to $\rho_0$
\cite{Hebeler:2010jx,Hebeler:2013nza}, even smaller radii can be
obtained. In particular, if the maximum mass is fixed to
$2\,M_{\odot}$, a $1.4 \, M_{\odot}$ star can have a radius $R_{1.4}$
down to about 10 km, while if the maximum mass is $2.4\,M_{\odot}$ then
$R_{1.4} \sim 11.5$ km. However, how to justify within a microscopic
calculation the needed polytropic EoS still needs to be clarified.

This seems to put a theoretical bias against the existence of stars
having very small radii. No single EoS exists at the moment which is
able to provide at the same time large masses for a few CSs and small
radii for others. Since the situation from the observational viewpoint
is still rather open, in this Letter we discuss a model which
satisfies those two conditions \footnote{A few EoSs provide radii of
  the order of 10 km for the maximum mass configuration
  \cite{Vidana:2010ip,Chamel:2012ea}.  If radii of the order of 10 km
  are not too rare, this does not solve the problem since stars having
  a very large mass should be rare, based on stellar evolution models
  and on SN simulations \cite{Ugliano:2012kq}.}. It is difficult to
have a unique family of CSs allowing both very small radii and very
massive configurations because to have small radii the EoS needs to be
rather soft. Therefore, large densities are reached in the center of
very compact stars, typically of the order of $5\div6\,\rho_0$ or
larger. On the other hand, to have very massive configurations the EoS
should be stiff at those same densities. No microscopic mechanism
exists to allow a sudden stiffening of the EoS at those large
densities. What we discuss in this Letter is instead a solution based
on {\it{two}} families of CSs, one made of hadrons and the other made
of deconfined quarks, QSs (we assume that the Bodmer-Witten hypothesis
to hold true). While in the literature many papers exist in which two
families have been discussed \cite{Berezhiani:2002ks,Bombaci:2004mt},
none takes into account the two constraints discussed above.

\begin{figure}[ptb]
\vskip 0.5cm
\begin{centering}
\epsfig{file=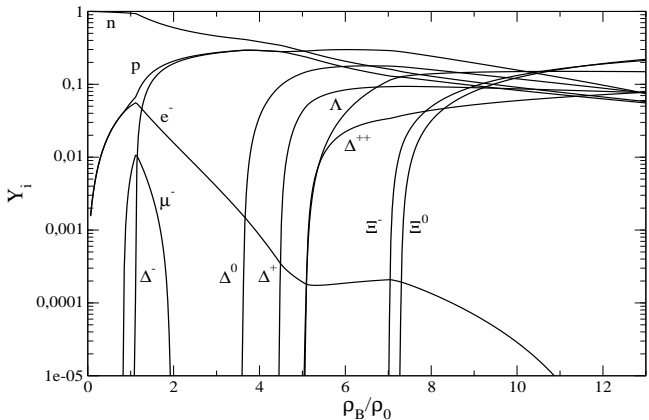,height=8.5cm,width=5.5cm,angle=-90}
\caption{Particles fractions as functions of baryon density, for $x_{\sigma\Delta}=1.25$, $x_{\omega\Delta}=1$}
\end{centering}
\end{figure}

It is rather natural to imagine that CSs with small radii are composed
of hadrons. As already mentioned above, at densities larger than about
$2.5\div 3\,\rho_0$ hyperons start appearing and in principle also
$\Delta(1232)$-resonances can be produced \footnote{The appearance at
  large densities of a mixture of nucleons and of $\Delta$s is natural
  from the viewpoint of the large-N$_c$ limit, in which nucleons and
  $\Delta$s are degenerate. This mixture could be the doorway towards
  the quarkionic phase, also speculated in the large-N$_c$ limit
  \cite{McLerran:2007qj}.}.  The production of these particles softens
the EoS and allows very compact configurations. On the other hand,
this same softening forbids this hadronic family of CSs to reach very
large masses
\cite{Baldo:1999rq,Vidana:2000ew,Massot:2012pf,Huber:1997mg,Schurhoff:2010ph}. It
is therefore very tempting to imagine that the most massive stars
correspond to QSs, since quark matter is known to be rather stiff and to support massive configurations
\cite{Fraga:2001id,Alford:2006vz,Kurkela:2009gj}. A crucial question
concerns the stability of the stars populating the hadronic branch:
when hyperons start being produced in the center of the star it is
relatively easy to have a transition to the more stable QS
configuration because droplets of strange quark matter can be formed.
For instance, the extremely compact hyperonic stars obtained
in Ref.\cite{SchaffnerBielich:2002ki}, would be unstable against decay into quark stars.
In order to have stable stars with very small radii we resort to the
production of $\Delta$ resonances which can shift the strangeness
production (hyperons) to higher densities.


In relativistic heavy ion collisions, where large values of
temperature and density can be reached, a state of resonance matter
may be formed and the $\Delta$s are expected to play a central role.
\cite{Zabrodin:2009fz,Hofmann:1994gn,Bass:1998vz,Lavagno:2010ah}.
\footnote{Within the non-linear Walecka model, it has been predicted
  that a phase transition from nucleonic matter to $\Delta$-excited
  nuclear matter can take place but the occurrence of this transition
  sensibly depends on the $\Delta$-meson coupling constants
  \cite{Li:1997yh,Kosov:1998gp,Lavagno:2010ah,Lavagno:2012bn}.}
Moreover, it has been pointed out that the existence of $\Delta$s can
be very relevant also in the core of neutron stars
\cite{Huber:1997mg,Xiang:2003qz,Chen:2007kxa,Chen:2009am,Schurhoff:2010ph}.

Concerning the hadronic EoS, we use the relativistic mean field model
with the inclusion of the octet of lightest baryons (nucleons and
hyperons) in the framework of the GM3 non-linear Walecka type model of
Glendenning-Moszkowsky \cite{Glendenning:1991es}. The values of the
meson-hyperon coupling constants have been fitted to reproduce the
potential depth of hyperons at saturation ($U_\Lambda^{N}=-28$ MeV,
$U_\Sigma^{N}=+30$ MeV, $U_\Xi^{N}=-18$ MeV)
\cite{Schaffner:1993nn,Schaffner:1995th}.  To incorporate
$\Delta$-isobars in the framework of effective hadron field theories,
a formalism was developed to treat $\Delta$ analogously to the
nucleon, taking only the on-shell $\Delta$s into account and the mass
of the $\Delta$s are substituted by the effective one in the mean
field approximation \cite{Boguta1982251,deJong:1992wm}.  The
Lagrangian density of the $\Delta$-isobars can then be expressed as
\cite{Boguta1982251,Li:1997yh,Kosov:1998gp}
\begin{eqnarray}
{\mathcal L}_\Delta=\overline{\psi}_{\Delta\,\nu}\, [i\gamma_\mu
\partial^\mu -(M_\Delta-g_{\sigma\Delta}
\sigma)-g_{\omega\Delta}\gamma_\mu\omega^\mu
 ]\psi_{\Delta}^\nu \, ,
\end{eqnarray}
where $\psi_\Delta^\nu$ is the Rarita-Schwinger spinor for the
$\Delta$-baryon.

Due to the uncertainty on the meson-$\Delta$ coupling
constants, we limit ourselves to considering only the couplings with
$\sigma$ and $\omega$ meson fields, which are explored in the
literature \cite{Li:1997yh,Kosov:1998gp,Jin:1994vw}.
\begin{figure}[ptb]
\vskip 0.5cm
\begin{centering}
\epsfig{file=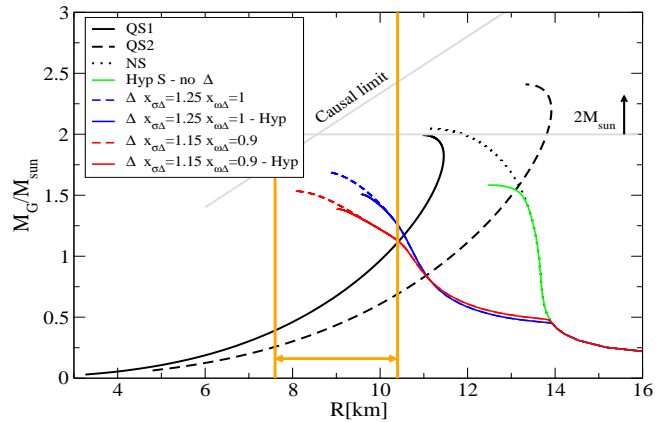,height=8.5cm,width=5.5cm,angle=-90}
\caption{Mass-radius relations of QSs (set1 and set2) and of hadronic stars.
The two orange lines correspond to the $1\sigma$ limit for the radii extracted from
the analyses of quiescent low-mass X-ray binaries \cite{Guillot:2013wu}.}
\end{centering}
\end{figure}
If the SU(6) symmetry is exact, one adopts the universal couplings
$x_{\sigma\Delta}=g_{\sigma\Delta}/g_{\sigma N}=1$ and
$x_{\omega\Delta}=g_{\omega\Delta}/g_{\omega N}=1$. However, the SU(6)
symmetry is not exactly fulfilled and one may assume the scalar
coupling ratio $x_{\sigma\Delta}>1$ with a value close to the mass
ratio of the $\Delta$ and the nucleon \cite{Kosov:1998gp}. On the
other hand, QCD finite-density sum rule results show that the Lorentz
vector self-energy for the $\Delta$ is significantly smaller than the
nucleon vector self-energy implying therefore $x_{\omega\Delta}<1$
\cite{Jin:1994vw}.   

In this paper we adopt two different choices for the $\Delta$-meson
couplings ($x_{\sigma\Delta}=1.25$, $x_{\omega\Delta}=1$ and
$x_{\sigma\Delta}=1.15$, $x_{\omega\Delta}=0.9$). Both
parameterizations are consistent with the experimental flow data of
heavy-ion collisions at intermediate energies
\cite{Danielewicz:2002pu}. Larger net attraction for $\Delta$-isobar
can imply mechanical instabilities in the EoS and this condition will
be explored in detail in future investigations. In Fig.~1 we display
the baryon density dependence of the particle's fractions.  It is
remarkable that the early appearance of $\Delta$ resonances, the first
one being the $\Delta^-$, considerably shifts the onset of hyperons
which start to form at densities of $\sim 5 \,\rho_0$ (see the curve
for the $\Lambda$'s).

A final comment concerning the experimental constraints on the density
dependence of the symmetry energy is in order \cite{Lattimer:2012xj}.
Within the GM3 parametrization here adopted, only the experimental
value of the symmetry energy at saturation $S_v$ is used ($S_v =32.5$
MeV in GM3) to fix the coupling between the $\rho$ meson and the
nucleons.  However, as shown in Ref. \cite{Lattimer:2012xj}, a remarkable
concordance among experimental, theoretical, and observational studies
has been found which allows to significantly constrain also the value
of $L$ (the derivative with respect to the density of the symmetry
energy at saturation). Extensions of the GM relativistic mean-field
model have been implemented which include $\rho$ meson
self-interaction terms.  These new parametrizations modify the density
dependence of the symmetry energy at supranuclear densities
\cite{Steiner:2004fi} and satisfy all the experimental constraints. It
turns out that, for pure nucleonic stars, $R_{1.4} \sim 12$ km (see for
instance \cite{Steiner:2012rk}), significantly smaller than the GM3
result: this is due to the fact that the more refined model provides a
softer EoS mainly in the density range $(1-2)\rho_0$.  A softening of
the EoS implies a delayed appearance of $\Delta$-resonances and
hyperons. The main aim of our work is to provide examples of hadronic
EoSs allowing for extremely compact stars with radii smaller than
10 km, what cannot be achieved by using pure nucleonic EoSs.  While it
is mandatory for our future studies to update the hadronic model in
order to take into account the symmetry energy experimental
constraints, on the other hand the formation of $\Delta$'s should
still be possible, although at larger densities.

For the quark matter EoS we rely, as is customary, to the simple MIT bag
model description in which confinement is provided by a bag constant
$B_{\mathrm{eff}}$ and the perturbative QCD interactions are effectively
included in the coefficient $a_4$ \cite{Alford:2004pf}. The total thermodynamical potential reads \cite{Weissenborn:2011qu}:
\begin{equation}
\Omega = \sum_{u,d,s,e} \Omega_i +\frac{3 \mu^4}{4\pi^2}(1-a_4)+B_{\mathrm{eff}}
\end{equation}
where $\mu$ is the quark chemical potential and
$\Omega_i$ are the thermodynamical potentials for non-interacting up, down,
and strange quarks and electrons.
The mass of the strange quark is fixed to $100$ MeV while the up and down quarks
are considered as massless. As shown in
Refs.~\cite{Weissenborn:2011qu}, in this
scheme it is possible to obtain stellar configurations up to two solar
masses or heavier. Here we will use the following
parameters sets: $B_{\mathrm{eff}}^{1/4}=142$ MeV -- $a_4=0.9$ (set1),
and $B_{\mathrm{eff}}^{1/4}=127$
MeV -- $a_4=0.6$ (set2) both taken from \cite{Weissenborn:2011qu}.
Set1 allows a maximum
mass for QSs of $2\, M_{\odot}$,
set2 has
been implemented to give an example of quark EoS for which the maximum
mass reaches $2.4 \,M_{\odot}$.

The mass-radius relations for QSs are displayed in Fig. 2 together
with hadronic stars.  The maximum mass of hadronic stars, containing
both $\Delta$ resonances and hyperons, is close to $1.5 \,M_{\odot}$
for the parameters' sets considered here.  When excluding hyperons and
$\Delta$-resonances the maximum mass of neutron stars reaches instead
a value of $\sim 2\, M_{\odot}$ but with a large radius. The
appearance of $\Delta$ resonances is crucial to obtain very compact
stellar configurations (as also shown in Ref. \cite{Schurhoff:2010ph})
with radii down to $8$ km (see red dashed line): the corresponding
mass-radius curves enter the area, framed by the two orange lines, of
very compact objects inferred in Ref. \cite{Guillot:2013wu}.  The
appearance of hyperons in the stars provides a further softening of
the EoS, reducing the maximum mass of $\sim 0.1\div 0.2 \,M_{\odot}$
(see solid/dashed red and blue lines).  On the other hand, the mass of
QSs can reach values compatible with the recent limit of $2
\,M_{\odot}$ (black solid line) or even higher values (black dashed
line). Notice that QSs mass-radius relations also enter the area of
very compact objects but for masses $\lesssim 1\, M_{\odot}$: such
light stars are difficult to produce in standard supernova simulations
and moreover the lightest known neutron star has a mass of $\sim 1.2
\,M_{\odot}$. The interpretation we propose here is that massive
stars, $M \gtrsim 1.5\div 1.6 \,M_{\odot}$, are QSs with radii $R
\gtrsim 11$ km whereas stars with $R \lesssim 10$ km are composed
mainly of nucleons and $\Delta$ resonances, with a maximum mass of
$\sim 1.5\div 1.6 \,M_{\odot}$. The tension between measurements of
large masses and small radii could be strengthened if a neutron star
more massive than $2 \,M_{\odot}$ is discovered favoring our
interpretation of two coexisting families of CSs (a possible candidate
is PSR B1957+20 with an estimated mass of $2.4\, M_{\odot}$
\cite{vanKerkwijk:2010mt}).

A crucial question concerns the astrophysical scenarios in which
hadronic and QSs are formed and how QSs can generate from hadronic
stars. In Fig.~3 we display the gravitational and baryonic masses as
functions of the radius for hadronic stars and QSs. On this plot it is
possible to construct a path for the formation of QSs from cold
hadronic stars accreting matter from a companion. The stellar
configuration labelled with B on the solid red line represents the
hadronic star for which hyperons start to form in the inner core
(notice that at the corresponding point on the baryonic mass curves,
the branch with hyperons deviates from the branch with only $\Delta$
resonances). The larger the mass of the star the larger its hyperon
content.  Notice that: i) only in the presence of hyperons, which
carry strangeness, can droplets of strange quark matter form via
nucleation \cite{Iida:1998pi}; ii) the star can ``decay'' into a QS with the same
baryonic mass since this process is energetically favored because the
gravitational mass of the configuration D is smaller than the one of
B. The energy released in the conversion of a hadronic star into a
quark star has been estimated in many papers and can easily reach
10$^{53}$ erg
\cite{Berezhiani:2002ks,Drago:2004vu,Bombaci:2004mt,Bombaci:2006cs}.

All the hadronic stellar configurations between B and A
can transform into QSs, the probability and velocity of
conversion depending on the specific microphysics process of formation
of the first droplets of quark matter and on the subsequent expansion of
the newly formed phase.
\begin{figure}[ptb]
\vskip 0.5cm
\begin{centering}
\epsfig{file=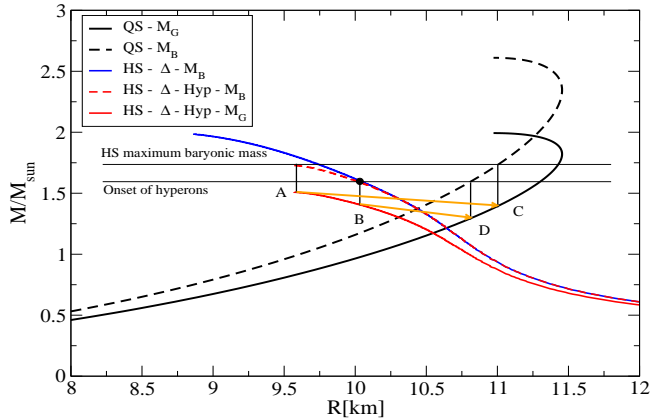,height=8.5cm,width=5.5cm,angle=-90}
\caption{Gravitational and baryonic mass-radius relations of QSs (set1) and
of hadronic stars (with and without hyperons, for $x_{\sigma\Delta}=1.25$, $x_{\omega\Delta}=1$).
A is the maximum mass
of hadronic stars containing hyperons. B is the gravitational mass
for which hyperons start to form.
The dot on the red dashed curve stands for the baryonic mass of B.
The quark stellar configurations
D and C have the same baryonic masses of B and A but smaller gravitational masses.}
\end{centering}
\end{figure}
There are many studies in the literature addressing these issues. In the
scenario here discussed, conversion of cold hadronic stars, quantum
nucleation represents a possible mechanism for the formation of the
first quark matter droplet
\cite{Iida:1998pi,Berezhiani:2002ks,Drago:2004vu,Bombaci:2004mt,Bombaci:2006cs}.
Once a seed of quark matter is formed, the conversion of the whole
hadronic star proceeds very fast, with time scales of the order of ms,
due to the development of hydrodynamical instabilities
\cite{Drago:2005yj,Herzog:2011sn,Pagliara:2013tza}. A detailed study
of the conversion process with the new proposed EoSs is mandatory for
future works.

Another scenario for the formation of quark stars is related to the
supernova explosion of massive progenitors.  Large densities can be
reached at the moment of the collapse, soon after the bounce, due to
the large fallback and hyperons can already appear at this stage,
immediately triggering the formation of quark matter.  There the
energy released in the conversion can help Supernovae to explode
\cite{Drago:2004vu,Drago:2008tb}.  In general the conversion of a
hadronic star into a QS will produce spectacular transient events such
as neutrino and gamma-ray-bursts.

There are many possible observables which could be used to test our
proposal in which most of the known neutron stars (with masses close
to $\sim 1.4 \, M_{\odot}$) are hadronic stars while massive stars are
more likely QSs (bare or with a crust).  We predict that massive CSs
also have large radii and, being composed by a different type of
matter with respect to the $1.4 \, M_{\odot}$ stars (in particular
regarding strangeness), should show anomalous cooling histories and
spinning frequency distributions; for instance, the
photon emission from the surface of a bare QS is very different from 
the one of neutron stars \cite{Harko:2004ts,Jaikumar:2004zy}. 
Moreover also quasi-periodic
oscillations of very massive CSs should differ from the ones of
hadronic stars \cite{Watts:2006hk}.

Finally let us discuss a well known argument against the
  coexistence of QSs and neutron stars, based on the production of
  strangelets during the merging of two CSs
  \cite{Friedman:1990qz,Madsen:2004ef}.  If at least one of the two
  CSs is a QS it is possible that strangelets are emitted polluting
  the whole Galaxy and triggering the conversion of all CSs into
  QSs. However recent numerical simulations of QSs' mergers have shown
  that, in many cases, a prompt collapse to a black hole occurs and no
  matter is ejected.  In particular, this occurs for values of the
  total mass of the merger larger than $\sim 3 M_{\odot} $ \cite{Bauswein:2008gx}.  It is
  clear that in the scenario here proposed this request is easily
  satisfied since for us QSs have masses larger than $\sim 1.5
  M_{\odot}$. Another possibility to prevent the strangelets pollution is 
offered by the observation that the burning of a neutron star into a QS
is uncomplete (at least in hydrodynamical simulations \cite{Drago:2005yj,Herzog:2011sn,Pagliara:2013tza}): it is therefore possible 
that a thick layer of hadronic matter survives shielding the inner quark matter core and making it more 
difficult to release strangelets.

We thank Anna Watts for valuable discussions.  G.P. acknowledges
financial support from the Italian Ministry of Research through the
program \textquotedblleft Rita Levi Montalcini\textquotedblright.


\end{document}